\documentclass[preprintnumbers,superscriptaddress,endnote,nofootinbib,aps,prd,floatfix,twocolumn]{revtex4-2}

\usepackage[utf8x]{inputenc}     
\usepackage{enumerate}
\usepackage[normalem]{ulem}
\usepackage{epsfig} 
\usepackage{float} 
\usepackage[caption = false]{subfig}
\usepackage{amsmath,amssymb}
\usepackage{wasysym} 
\usepackage{mathrsfs} 
\usepackage{amsfonts} 
\usepackage{amsbsy} 
\usepackage{amscd} 
\usepackage{pstricks} 
\usepackage{multirow} 
\usepackage{tikz}
\usepackage{color}
\usepackage{slashed}
\usepackage{array}
\usepackage{tablefootnote}
\usepackage{csvsimple} 
\usetikzlibrary{arrows,positioning,shapes.geometric} 
\usepackage[compat=1.1.0]{tikz-feynman}         
\usepackage{slashed}
\usepackage{multirow}
\usepackage{tabularx}
\usepackage{soul}  
\usepackage{listings} 
\usepackage{hyperref}
\usepackage{mathtools}
\hypersetup{colorlinks=true, citecolor=blue, urlcolor=blue, linkcolor=blue}
\usepackage{color}

\usepackage{amsthm}
\newcommand{\ket}[1]{|#1\rangle}

\usepackage{amsthm}

\DeclareMathOperator\supp{supp}
\usepackage{mathtools}
\DeclarePairedDelimiter\ceil{\lceil}{\rceil}
\DeclarePairedDelimiter\floor{\lfloor}{\rfloor}

\usepackage{booktabs,enumitem,algorithm,algpseudocode}

\begin{document}
\title{Improved Ground State Estimation in Quantum Field Theories \\ via Normalising Flow-Assisted Neural Quantum States}

\begin{abstract}
We propose a hybrid variational framework that enhances Neural Quantum States (NQS) with a Normalising Flow-based sampler to improve the expressivity and trainability of quantum many-body wavefunctions. Our approach decouples the sampling task from the variational ansatz by learning a continuous flow model that targets a discretised, amplitude-supported subspace of the Hilbert space. This overcomes limitations of Markov Chain Monte Carlo (MCMC) and autoregressive methods, especially in regimes with long-range correlations and volume-law entanglement. Applied to the transverse-field Ising model with both short- and long-range interactions, our method achieves comparable ground state energy errors with state-of-the-art matrix product states and lower energies than autoregressive NQS. For systems up to 50 spins, we demonstrate high accuracy and robust convergence across a wide range of coupling strengths, including regimes where competing methods fail. Our results showcase the utility of flow-assisted sampling as a scalable tool for quantum simulation and offer a new approach toward learning expressive quantum states in high-dimensional Hilbert spaces.
\end{abstract}

\preprint{IPPP/25/33}
\author{Vishal S. Ngairangbam}\email{vishal.s.ngairangbam@durham.ac.uk}
\author{Michael~Spannowsky} \email{michael.spannowsky@durham.ac.uk}
\author{Timur~Sypchenko}\email{timur.sypchenko@durham.ac.uk}
\affiliation{Institute for Particle Physics Phenomenology, Durham University, Durham DH1 3LE, United Kingdom\\[0.1cm]}

\maketitle
\section{Introduction}

Quantum many-body systems lie at the heart of condensed matter physics, quantum field theory, and quantum information science. A central computational challenge in this domain is estimating ground states of strongly interacting quantum systems. These states encode essential information about phases of matter, criticality, and entanglement structure \cite{PhysRevB.14.1165,RevModPhys.80.517,PhysRevE.100.062134,Hayden:2007cs,PhysRevD.109.066010,PhysRevB.101.174312,PhysRevLett.125.130601,Lashkari2013,PhysRevA.94.040302,PhysRevA.102.022402,Criado:2021gzp,Criado:2023rcx,Schenk:2020lea,Araz:2022tbd,Lin:2024eiz,Lu:2024utm}. However, the exponential growth of Hilbert space with system size renders brute-force approaches intractable, and even variational and Monte Carlo techniques struggle in the presence of high entanglement or nonlocal interactions. Traditional methods, such as tensor networks, perform well in low-dimensional, weakly entangled regimes but often fail to generalise efficiently to volume-law entangled states or higher-dimensional systems. 

In recent years, Neural Quantum States (NQS) have emerged as a promising machine learning-based variational ansatz for quantum wave functions. Originally proposed by Carleo and Troyer \cite{doi:10.1126/science.aag2302}, NQS represent the amplitudes (and possibly phases) of quantum states using neural networks, enabling expressive, compact approximations of complex many-body wave functions. These models have demonstrated remarkable performance in capturing ground states of a variety of spin systems ~\cite{PhysRevLett.131.036502,PhysRevE.109.064123,PhysRevLett.134.079701,Chen2024,PhysRevB.110.205147,PhysRevB.110.205147,PhysRevResearch.5.033116,10.21468/SciPostPhys.10.6.147,Roca-Jerat:2024fud,Romero:2024eyz,Rende:2024bsn,Rende:2024ane,Joshi:2024mey,Beck:2024kxo,Duric:2024egn,Wu:2023mxn,Joshi:2023xdj,Mezera:2023scr} , especially when combined with variational Monte Carlo sampling \cite{Metropolis:1953am,hastings,RevModPhys.73.33,PhysRevResearch.5.033041}. Public frameworks such as NetKet~\cite{netket2:2019,10.21468/SciPostPhysCodeb.7,10.21468/SciPostPhysCodeb.7} have made NQS methods widely accessible, fostering rapid development of new architectures and optimisation techniques.

While the theoretical capacity of deep learning architectures to efficiently represent entangled quantum states is known~\cite{PhysRevLett.122.065301,Gao2017,Yang:2024yxu}, a major bottleneck persists: sampling from the variational distribution. The quality of the learned quantum state is fundamentally tied to the efficiency and fidelity of the sampling method used during training. Autoregressive models~\cite{JMLR:v17:16-272,PhysRevLett.124.020503,PhysRevResearch.5.013216,10.21468/SciPostPhys.14.6.171,PhysRevB.107.075147,PhysRevLett.128.090501} and Markov Chain Monte Carlo (MCMC) approaches, though popular, can suffer from inefficiencies~\cite{Bortone:2023arv} in high-dimensional or rugged energy landscapes—especially in systems with strong non-local correlations or topological constraints.

In this work, we propose a novel hybrid strategy that integrates normalising flows (NFs) \cite{pmlr-v37-rezende15} into the NQS framework to address this sampling challenge. NFs are a class of generative models that learn complex probability distributions by transforming a simple prior through a sequence of invertible, differentiable mappings. They allow for efficient and exact sampling and offer a natural pathway to model expressive distributions in large, structured spaces. Due to their inherent construction in smooth spaces, their usage in literature has so far been restricted to simulating quantum systems with continuous degrees of freedom~\cite{Stokes_2023,Lawrence:2024ebc}. Efforts in machine learning literature to adapt NFs to the discrete domain~\cite{NEURIPS2019_e046ede6} based on the composition of discrete transformations and straight-through gradient estimation~\cite{DBLP:journals/corr/BengioLC13} have met with success for low cardinality finite sample spaces. We discretise the posterior, not the latent space, by partitioning it into mutually exclusive regions $R_x$.  This removes any reliance on straight-through gradient estimators.
The continuous flow is mapped onto discrete basis states, giving a sampler that targets highly-entangled subspaces of an exponentially large Hilbert space, yet keeps the continuous parameter dimension linear in system size.

We demonstrate our approach on a prototypical spin system—a transverse-field Ising model with all-to-all interactions—and show that NF-enhanced sampling improves ground state energy estimation over standard variational techniques. Our method enables efficient exploration of nonlocal quantum configurations and opens the door to more expressive, sample-efficient learning of complex quantum states.

We find that augmenting Neural Quantum States with Normalising Flow-based sampling dramatically improves the estimation of ground states in quantum many-body systems, particularly in regimes with strong entanglement and non-local interactions. By decoupling the sampling task from the variational ansatz and leveraging continuous flows to target effective subspaces of the Hilbert space, our approach overcomes key limitations of Markov Chain Monte Carlo (MCMC) and autoregressive methods. Applied to the transverse-field Ising model with varying interaction ranges and strengths, our method consistently achieves lower ground state energy errors than matrix product states and conventional neural samplers. Crucially, it remains robust at larger system sizes where exact diagonalisation fails and competing methods collapse into trivial modes. These results suggest that flow-assisted subspace sampling can serve as a scalable and expressive foundation for quantum simulation.

In Section~\ref{sec:nqs}, we review the variational Monte Carlo framework for simulating quantum field theories using Neural Quantum States, and discuss the limitations imposed by conventional sampling algorithms. Section~\ref{sec:groundstate} introduces our core methodological contribution: a Normalising Flow-based sampler that decouples support discovery from amplitude learning, including the discretisation scheme and training strategy that enable sampling from highly entangled subspaces of Hilbert space. Section~\ref{sec:numerics} presents a detailed numerical analysis on the transverse-field Ising model, demonstrating the performance of our method across different system sizes, interaction strengths, and entanglement regimes, with comparisons to state-of-the-art baselines. We conclude in Section~\ref{sec:conclusion} with a summary of our results and an outlook on future directions for generative model-assisted quantum simulation.

\section{Simulating Quantum Field Theories with Neural Quantum States and Machine Learning}
\label{sec:nqs}

\subsection{Variational Monte Carlo} 
The numerical simulation of strongly interacting quantum field theories (QFTs) remains an outstanding computational problem in theoretical physics. In lattice regularisations of QFTs, one typically works with a discretised Hilbert space $\mathcal{H} \cong \mathbb{C}^{d^N}$, where d is the local Hilbert space dimension and N the number of degrees of freedom (e.g., lattice sites or modes). The exponential scaling of $\dim \mathcal{H}$ with $N$ renders exact diagonalization intractable beyond modest system sizes.
Let $\{ |\mathbf{x}\rangle \}$ denote an orthonormal basis in $\mathcal{H}$. The linear combination gives a generic pure quantum state:
\begin{equation}
|\psi\rangle = \sum_{\mathbf{x} \in \mathcal{B}} \psi(\mathbf{x}) |\mathbf{x}\rangle,
\end{equation}
where $\mathcal{B} \subset \{0, 1, \dots, d-1\}^N$ indexes the basis configurations, and $\psi(\mathbf{x}) \in \mathbb{C}$ are complex amplitudes.
The variational problem consists in finding a parametrised approximation $\psi_\theta(\mathbf{x}) \approx \psi_0(\mathbf{x})$, where $\psi_0$ is the ground state of a Hamiltonian $\hat{H}$. This ground state defines a probability mass function on the basis $\mathcal{B}$, typically exhibits non-trivial entanglement scaling and, generally, has significant support only on a relatively small subspace $\mathcal{H}_\mathrm{eff} \subset \mathcal{H}$ defined by a small $\epsilon>0$ as :

\begin{equation}
    \mathrm{supp}(\psi_0) = \left\{ \mathbf{x} \in \mathcal{B} \,\big|\, |\psi_0(\mathbf{x})|^2 > \varepsilon \right\},~ \dim \mathcal{H}_\mathrm{eff} \ll \dim \mathcal{H}.
\end{equation}

Thus, the variational task is governed by two interdependent challenges. First, one must identify the effective support of the ground state wave function, that is, determine the subset of basis vectors that contribute significantly to the ground state—this corresponds to discovering the relevant subspace $\mathcal{H}_\mathrm{eff} \subset \mathcal{H}$. Second, having identified this support, it is crucial to accurately optimize the amplitude and, where applicable, the phase of the wave function coefficients $\psi(\mathbf{x})$ for each contributing configuration $\mathbf{x} \in \mathcal{H}_{\mathrm{eff}}$. Both steps are essential to faithfully approximate the ground state in high-dimensional Hilbert spaces. 
Neural Quantum States address the second challenge by parametrising the amplitude and phase using a neural network

\begin{equation}
\psi_\theta(\mathbf{x}) = \sqrt{p_\theta(\mathbf{x})} \, e^{i \alpha_\theta(\mathbf{x})},
\end{equation}
where $p_\theta$ is a normalized probability distribution and $\alpha_\theta$ the phase function. The ground state energy is estimated via a variational minimisation of the energy expectation. 
\begin{equation}
\label{eq:var_energy}
E[\psi_\theta] = \frac{\sum_{\mathbf{x}, \mathbf{x}’} \psi_\theta^*(\mathbf{x}) \langle \mathbf{x} | \hat{H} | \mathbf{x}’ \rangle \psi_\theta(\mathbf{x}’)}{\sum_{\mathbf{x}} |\psi_\theta(\mathbf{x})|^2}.
\end{equation}

Often, however, notably not in the method we propose in Sec~\ref{sec:groundstate}, this expression is evaluated via Monte Carlo sampling using the variational distribution $p_\theta(\mathbf{x})$~\cite{netket2:2019}. The local energy at a given configuration is defined as
\begin{equation}
E_{\mathrm{loc}}(\mathbf{x}) = \sum_{\mathbf{x}’} \frac{\langle \mathbf{x} | \hat{H} | \mathbf{x}’ \rangle \psi_\theta(\mathbf{x}’)}{\psi_\theta(\mathbf{x})},
\end{equation}
and the variational gradient reads
\begin{equation}
\nabla_\theta E = 2\,\mathrm{Re}\left\langle \left(E_{\mathrm{loc}}(\mathbf{x}) - \langle E_{\mathrm{loc}} \rangle\right) \nabla_\theta \log \psi_\theta(\mathbf{x}) \right\rangle.
\end{equation}

\subsection{Bottlenecks in Sampling algorithms}
Although neural networks can theoretically represent entangled states with polynomially scaling number of parameters~~\cite{PhysRevLett.122.065301,Gao2017,Yang:2024yxu}, this is yet to be translated to practical utility.  This is due to the inefficiency of Monte-Carlo based sampling algorithms to efficiently sample from $\mathcal{H}_{\mathrm{eff}}$.  
In other words, even if $\psi_\theta$ has the capacity to approximate $\psi_0$, poor sampling will fail to expose $\mathcal{H}_\mathrm{eff}$, and the model may converge to local minima corresponding to energetically suboptimal states. 
 
 MCMC sampling, while widely used, tends to explore configuration space locally and can be inefficient in uncovering distant but important configurations, especially in systems with long-range correlations, topological sectors, or volume-law entanglement. This is due to the imposition of the sufficient, but not necessary \emph{microscopic reversibility} condition. If the target distribution has a very high mass at a particular point, the trajectory cannot escape the vicinity of such points and leads to suboptimal exploration of the state space. Additionally, the algorithm needs an exponential number of time steps if the relevant subset of $\mathcal{B}$ describing $\mathcal{H}_{\mathrm{eff}}$ is scattered over the entire set. This happens in highly entangled quantum states where information is shared across each constituent physical subsystem.  
 
 Autoregressive networks~\cite{JMLR:v17:16-272,PhysRevLett.124.020503} go beyond traditional MCMC sampling for approximating the true distribution where the probability distribution $p(\mathbf{x})$ is decomposed as a product of conditional distributions with the \emph{autoregressive property}. The decomposition of $p(\mathbf{x})$ into conditional probabilities between subsets of the argument $\mathbf{x}=(x_1,x_2,...,x_N)$ follows the autoregressive property if it makes no assumption of conditional independence, i.e. 
 \begin{equation*}
p(\mathbf{x})=\prod_{i=1}^{N}\; p_i(x_{i}|x_{i-1},x_{i-2},...,x_{1}) \quad.
 \end{equation*} 
 Theoretically, any probability distribution has such a decomposition, but not every conditional decomposition leads to an efficient representation of a distribution. Therefore, the expressivity is practically limited by the order of the chosen conditionals. Specifically for quantum systems, the $N$-dimensional local quantum number vector $\mathbf{x}$ must be decomposed into a specific order to construct the conditional distributions. This assumed order then restricts the nature of correlations that can be captured between the different subsystems' states that make up the vector $\mathbf{x}$. Heuristically, one expects that such an artificial decomposition would fail to efficiently capture long-range correlations between any two given subsystems in strongly entangled states.

\subsection{Decoupled Subspace Sampling via Generative Models}
From the discussions above, one of the bottlenecks in NQS simulation is efficient sampling of $\mathcal{H}_{\mathrm{eff}}$, especially when $\dim \mathcal{H}_{\mathrm{eff}}<< \dim\mathcal{H}$. Since in cases where $\dim \mathcal{H}_{\mathrm{eff}}\sim\dim\mathcal{H}$, one has to directly contend with the exponentially large Hilbert space, we primarily focus on the former. 
In such cases, our proposed method aims to leverage modern generative machine learning algorithms through which we can learn to sample from 
$\mathcal{H}_{\mathrm{eff}}$. 

\begin{center} 
\begin{figure*}[t]
  \centering
  \includegraphics[scale=0.5]{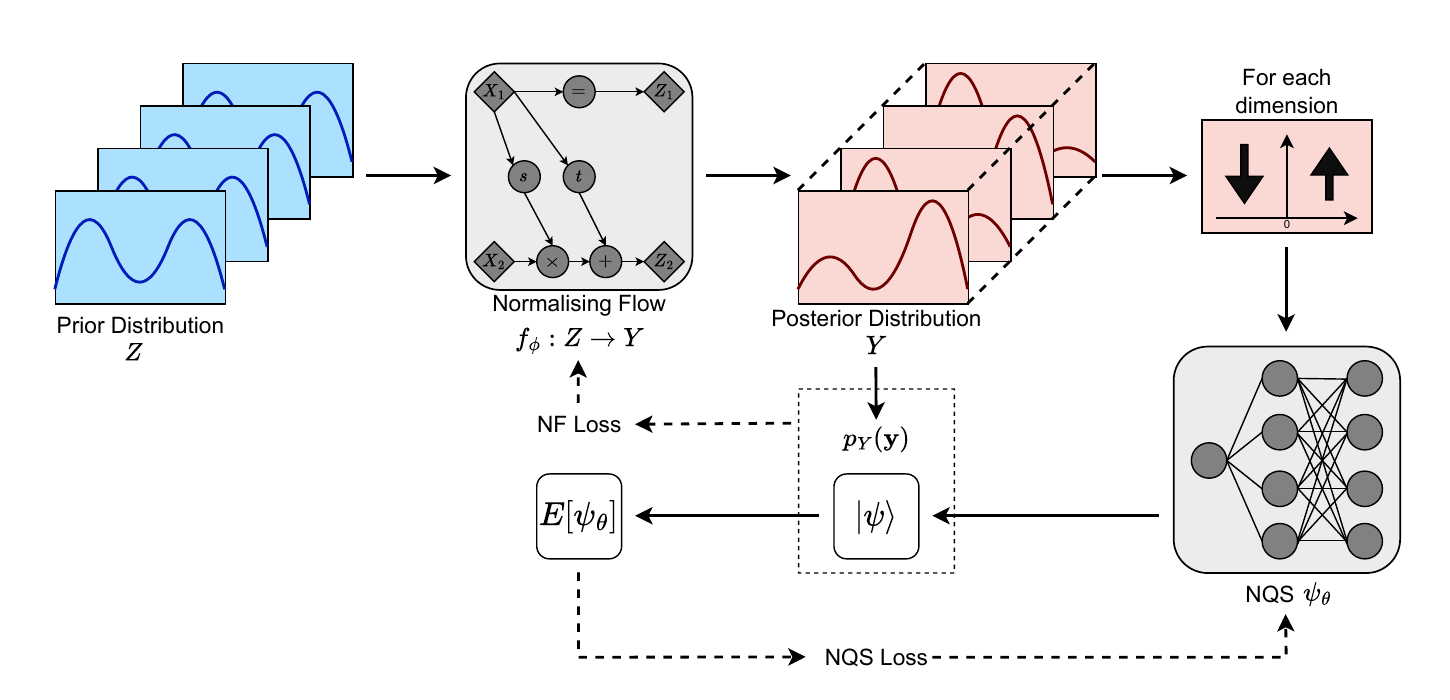}
  \caption{Schematic of the decoupled-subspace variational ansatz: a Normalising Flow (left) learns the support, while an independent NQS (right) learns amplitudes.
 Samples from a continuous prior distribution $p_Z(\mathbf{z})$ are transformed into posterior samples via the flow transformation $f_\phi:Z\to Y$. These samples are transformed into unique discrete basis states $\mathbf{x}$ based on fixed predetermined regions $R_\mathbf{x}$ in $Y$. One defines an independent Neural Quantum State $\psi_\theta$ on these samples via which one evaluates the variational energy $E[\psi_\theta]$ and the discrete target probability $p_\theta(\mathbf{x})$. This drives the learning of a continuous target probability $p_Y(\mathbf{y})$ via the normalising flows, which has modes in those regions belonging to basis states with high contribution to the ground state probability.}
  \label{fig:flow_nqs_pipeline}
\end{figure*}
\end{center}

Traditionally, MCMC methods relied on the variational wavefunction's probability estimate to sample new basis states. This has carried over to NQS literature including autoregressive methods since $\mathcal{H}_{\mathrm{eff}}$ is defined by $\supp(\psi_0)$. The main novelty of our work is offloading the task of sampling from $\mathcal{H}_{\mathrm{eff}}$ to a generative model with independent parameters from the variational ansatz $\psi_\theta$.  To do this, we need a generative model capable of representation learning with the ability to generate basis samples in $\mathcal{B}$. We choose normalising flows from other choices of representation-learning capable algorithms like Variational Autoencoders and Generative Adversarial Networks due to their additional ability to estimate the likelihood explicitly.

A schematic representation of the decoupled subspace sampling Neural Variational Ansatz using normalising flows is shown in Fig.~\ref {fig:flow_nqs_pipeline}. By sampling from the prior distribution, the normalising flow model outputs continuous posterior samples in $Y$. With a discretisation scheme, the posterior samples are converted to basis states describing a closed subspace of $\mathcal{H}$. One then constructs a variational wave function $\ket{\psi_\theta}$ in this subspace via an independent NQS model. The probabilities obtained via $\ket{\psi_\theta}$ drive the learning of the normalising flows model dependent on the variational energy $E[\psi_\theta]$. While the NQS model also learns the coefficients based on the variational energy, its primary role is to drive the generative model's learning. Once the generative model has converged, one can choose from the plethora of available deep learning algorithms with no additional overhead for the sampler learning process.  

NFs can therefore be harnessed to improve the variational sampling strategy. Instead of relying on (say) local MCMC updates to draw configurations $\mathbf{x}$ from the variational distribution $p_\theta(\mathbf{x})$ obtained from an independently parametrised NQS $\psi_\theta$, one can learn a flow $f_\phi$ that directly maps latent samples $\mathbf{z} \sim p_Z$ to the support of the quantum state via an independent discrete distribution $p_\phi(\mathbf{x})$ that need not exactly match $p_\theta$. This decouples the sampling process from locality constraints and allows the model to efficiently explore disconnected, non-local regions of configuration space. Such capability is especially critical when the ground state lies in a complex subspace $\mathcal{H}_\mathrm{eff}$ with high entanglement or non-trivial topological structure. Note, however, that we will need to rely on the variational energy estimate to tune both the NF model and the NQS model.  

Crucially, normalising flows contribute to both aspects of the ground state learning problem. First, by enabling non-local, differentiable sampling, they significantly enhance the discovery of the effective support $\mathcal{H}_\mathrm{eff}$, even in rugged or multi-modal configuration spaces. Second, since the sampling process is tightly coupled to the variational loss, the flow parameters are concurrently optimised to allocate probability mass accurately, thereby improving the amplitude estimation of the wave function. Together, these capabilities make normalising flows a powerful addition to the NQS toolbox, particularly for quantum systems characterised by long-range interactions, volume-law entanglement, or intricate phase structure.

\section{Learning to sample the ground state subspace}
\label{sec:groundstate}

The primary disadvantage of both MCMC sampling and autoregressive methods arises from their artificial imposition of sequentiality, albeit in different ways. The former sequentially explores local regions in the state space even though the closed subspace $\mathcal{H}_{\mathrm{eff}}$ is described by an unordered subset of $\mathcal{B}$.  Meanwhile, the latter assumes a sequential decomposition of the individual subsystems, which works for short-range and weakly entangled subsystems but will become increasingly inefficient for strongly correlated states. Thus, under the assumption that the initial distribution of the decoupled NF model is uniform in $\mathcal{H}$, and given its distributional universality~\cite{nf_universal}, our decoupled sampling approach would not be limited by such factors.  

\subsection{Normalizing Flows}

NFs are a class of generative models that transform a simple base distribution into a complex target distribution through a sequence of invertible and differentiable mappings. Mathematically, let $\mathbf{z} \in \mathbb{R}^n$ be a sample from a known prior distribution $p_Z(\mathbf{z})$, such as a standard Gaussian. A normalizing flow defines an invertible transformation $f_\phi : \mathbb{R}^n \rightarrow \mathbb{R}^n$, parametrized by $\phi$, which maps $\mathbf{z}$ to a new variable $\mathbf{y} = f_\phi(\mathbf{z})$. The resulting probability density over $\mathbf{y}$ is given by the change-of-variables formula
\begin{equation}
p_Y(\mathbf{y}) = p_Z(f_\phi^{-1}(\mathbf{y})) \left| \det \left( \frac{\partial f_\phi^{-1}}{\partial \mathbf{y}} \right) \right|.
\end{equation}
The function $f_\phi$ generally consists of multiple diffeomorphisms $f_\phi = f_K \circ \cdots \circ f_1$, where each constituent has the same form that together have enough expressive power while allowing efficient Jacobian evaluations. In practice, one evaluates the log probabilities as
\begin{equation}
\label{eq:post_y_prob} 
\log p_Y(\mathbf{y})=\log p_Z(\mathbf{z})-\sum_{i=1}^K\log\left| \det \left(\frac{\partial f_i}{\partial \mathbf{z}_i}\right)\right|\quad, 
\end{equation} 
where in the second term we define $\mathbf{z}_0\equiv \mathbf{z}$, $\mathbf{y}\equiv\mathbf{z}_K$, and $\mathbf{z}_{i}=f_i(\mathbf{z}_{i-1})$ for $K\geq i>0$.

They can be used for (i) generative tasks where one has a direct or indirect measure of the target distribution but it is generally difficult to sample from, and (ii) maximum likelihood estimation where one has samples from the target distribution but does not have explicit knowledge of the underlying probability. For the first case, the parameters $\phi$ are trained to match a target distribution by minimising a divergence measure—commonly the Kullback-Leibler divergence while for the second, one utilises the negative log likelihood for maximum likelihood estimation (MLE). Clearly, our aim is to learn to sample from a target distribution where we have the energy expectation as an indirect measure for the probability mass on $\mathcal{B}$. However, due to the discrete nature of our target space and the continuous nature of the normalising flow's posterior space, we utilise a hybrid structure where we rely on maximum likelihood estimation in independent regions that smooth the gradients of $\phi$.

\subsection{Discretising the Posterior Space}
To apply flows in discrete quantum settings where the Hilbert space is spanned by a finite basis $\{ |\mathbf{x}\rangle \}$, a discretisation scheme is introduced. One defines mutually disjoint regions $\{ R_{\mathbf{x}} \}$ in the continuous posterior space of the flow such that each configuration $\mathbf{x}$ is associated with a region $R_{\mathbf{x}} \subset \mathbb{R}^n$. The probability assigned to $\mathbf{x}$ by the flow model is then given by the integral
\begin{equation}
\label{eq:discretization}
p_\phi(\mathbf{x}) = \int_{R_{\mathbf{x}}} p_Y(\mathbf{y}) \, d\mathbf{y}.
\end{equation}

We approximate this integral by Monte Carlo sampling inside each region \(R_{\mathbf{x}}\).  Let \(M_{\mathbf{x}}\) be a set of $N_{\mathrm{MC}}$ points sampled in \(R_{\mathbf{x}}\). Then
\begin{equation}
\label{eq:mc_estimate}
\hat p_\phi(\mathbf{x})
= \frac{\mathrm{Vol}\bigl(R_{\mathbf{x}}\bigr)}{\lvert M_{\mathbf{x}}\rvert}
  \sum_{\mathbf{y}\in M_{\mathbf{x}}} p_Y(\mathbf{y})
\;\approx\;
  \int_{R_{\mathbf{x}}} p_Y(\mathbf{y})\,d\mathbf{y}\,.
\end{equation}

This construction enables the use of continuous flows to represent and sample from discrete variational distributions over Hilbert space basis elements. During training, the parameters of the flow are adjusted to minimise the discrepancy between the model distribution $p_\theta(\mathbf{x})$ and the desired probability $|\psi_0(\mathbf{x})|^2$, indirectly driving the variational wave function toward the true ground state.

We expect the flow to concentrate its probability density in exactly those regions $R_{\mathbf{x}}$ whose corresponding basis states $\mathbf{x}$ carry high ground‐state weight $|\psi_0(\mathbf{x})|^2$. As stated before, we designate the task of precisely estimating the amplitude to an independent NQS model. Therefore, each integral weight $p_\theta(\mathbf{x})$ needs not be very close to the exact value of $|\psi_0(\mathbf{x})|^2$ as long as an i.i.d sampling in the prior with $p_Z(\mathbf{z})$ maximally produces discrete basis states describing the subspace $\mathcal{H}_{\mathrm{eff}}$. 

We discretise the flow’s continuous output by mapping each coordinate to its nearest basis state. In the two-level (spin-$\frac{1}{2}$) case via 
\begin{equation}\label{eq:sign_function}
s_i = \mathrm{sign}(y_i)\quad(\text{up/down}).
\end{equation}
For a general $d$‐dimensional local sites, one arranges $d$ outputs per site, applies a softmax, and selects the class with highest probability (argmax).  The total latent dimension $n = N\times d$ still grows only linearly in system size, yet this softmax‐based scheme can represent all $d^N$ discrete configurations efficiently.

\subsection{Training Strategy}

The presented algorithm comprises two modules: the normalising flows model $f_\phi:\mathbb{R}^n\to\mathbb{R}^n$ that learns to sample from $\mathcal{H}_{\mathrm{eff}}$, and an independently parametrized NQS model $\psi_\theta$, that drives the learning of $p_\phi(\mathbf{x})$ via $f_\phi$. The aim of $\psi_\theta$ is to drive the learning of the NF model rather than to exactly learn the ground state coefficients during training. Once the NF model is trained to reproduce basis states in $\mathcal{H}_{\mathrm{eff}}$, one need not rely on $\psi_\theta$ and can, in principle, train various deep learning algorithms with little overhead in the sampling procedure. 

For learning to sample maximally from $\mathcal{H}_{\mathrm{eff}}$, we want to tune the parameters of the flow $f_\phi:\mathbb{R}^n\to\mathbb{R}^n$ so that the continuous posterior's modes will be located in regions $R_\mathbf{x}$ belonging to $\mathcal{H}_{\mathrm{eff}}$. The initial discretised posterior should be uniform on $\mathcal{B}$ to allow for a good convergence to the correct subspace. This can be achieved by making the prior have equal integrated weight in the regions $R_\mathbf{x}$ and then making the flow transformation approximately follow an identity map.

The whole algorithm goes as follows: 
\begin{enumerate} 
\item Sampling from the prior distribution $p_Z(\mathbf{z})$, we get discretized posteriors $\mathbf{x}$ via the continuous posterior $p_Y(\mathbf{y})$. 
\item Since it is not guaranteed that the batch contains only distinct basis states, we keep single unique values of $\mathbf{x}$, i.e. the discrete samples $S=\{\mathbf{x}\}$ is a set. 
\item For basis states $\mathbf{x}$ in $S$, we evaluate the discrete target probability of the NF model from the NQS model, i.e. we define $p_\theta(\mathbf{x})=|\psi_\theta(\mathbf{x})|^2/\sum_{\mathbf{x}'} |\psi_\theta(\mathbf{x'})|^2$. 
\item In each region $R_\mathbf{x}$ that is contained in $S$, we independently sample  $N_{\mathrm{MC}}$ values of $\mathbf{y}\in R_{\mathbf{x}}$ from a multivariate normal distributions with a scalar covariance matrix such that the mean is located at least five standard deviations away from any adjacent regions.
As long as we keep $N_{\mathrm{MC}}$ not too large, it is highly unlikely for any set $M_{\mathbf{x}}$ to contain samples from a region outside $R_\mathbf{x}$ and even more unlikely that the sample mean over $M_\mathbf{x}$ moves outside of $R_\mathbf{x}$. 
\item The continuous target probability of the Normalising Flow's posterior is defined as the $p_\theta(\mathbf{x})$ weighted sum of the independent normal distributions in each region $R_\mathbf{x}$.

\end{enumerate} 
Thus, within such a procedure, the NQS energy expectation $E[\psi_\theta]$ can drive the learningof the NF's parameters. 
We define the loss function for the NF parameters as

\begin{equation}
\label{eq:nf_loss_mc}
\mathcal{L}_\phi[p_Y,S]
= -\frac{\lvert E[\psi_\theta]\rvert}{\lvert S\rvert}
  \sum_{\mathbf{x}\in S} p_\theta(\mathbf{x})\,
  \log\!\bigl(\hat p_\phi(\mathbf{x})\bigr)\,,
\end{equation}
where \(\hat p_\phi(\mathbf{x})\) is the Monte Carlo estimator defined in Eq.~\ref{eq:mc_estimate}.

 Here, the parameters $\theta$ that enter $p_\theta(\mathbf{x})$ and $E[\psi_\theta]$ are kept fixed, i.e. during gradient evaluation of $\mathcal{L}_\phi$, we evaluate the partial derivatives with respect to $\phi$ alone. Looking at the expression without $|E[\psi_\theta]|$, $\mathcal{L}_\phi$ can be interpreted as a cross-entropy loss between the discrete target distribution and the Monte Carlo estimate of the continuous target probabilities within each region $R_\mathbf{x}$. The multiplicative factor of the energy $|E[\psi_\theta]|$ prioritises those directions with higher energy magnitude, since our systems have negative ground state energies. We find that such a modification helps in better convergence. 
The loss function used to update the parameters $\theta$ is taken directly as the energy expectation $E[\psi_\theta]$. We implement a gradient clipping threshold of 1000, as due to larger values of the energy expectation at large $N$, the gradients at the last layer of the NQS become unstable.
We update the parameters of both models over a batch of $B=\{S_1,S_2,...,S_{|B|}\}$ via the averaged loss function 
\begin{equation}
\mathcal{L}=\frac{1}{|B|}\sum_{i} \left(\mathcal{L}_\phi[p_Y(\mathbf{y}),S_i]+E[\psi_\theta]\right)~.
\end{equation}

\section{Numerical Analysis}
\label{sec:numerics}

\subsection{Non-local Transverse Field Ising Model}
We study the proposed algorithm using the prototypical transverse field Ising model~\cite{PFEUTY197079} in 1D with periodic boundary conditions and differing levels of connections and interaction strengths between the sites. For a given undirected graph $\mathcal{G}(\mathcal{N},\mathcal{E})$ with nodes $\mathcal{N}$ representing local sites, and edges $\mathcal{E}$ representing interactions, we have the Hamiltonian
\begin{equation} 
    H = -V \sum_{\{i,j\}\in\mathcal{E}} \sigma_i^z \sigma_j^z - \sum_{i\in\mathcal{N}} \sigma_i^x\quad.
    \label{eq:transverse_field_ising_model}
\end{equation} 
As a proof-of-principle, we first consider exactly solvable systems with $N\in\{10,15,20\}$. 

For each \(N\), we impose periodic boundary conditions (PBC) and study three interaction lengths \(L\): 
\begin{enumerate}[label=(\roman*)]
  \item Nearest neighbor coupling, \(L=1\).
  \item Intermediate range, \(L=\bigl\lfloor N/4\bigr\rfloor\).
  \item Maximal range under PBC, \(L=\bigl\lfloor N/2\bigr\rfloor\).
\end{enumerate}
For the first two cases, we set $V=1$, while for the latter we study the algorithmic performance of various methods by varying the interaction strength $V\in\{0.1,0.5,1.0\}$,
which indirectly increases the probability weight on the fully aligned (all-up/all-down) configurations, making it more challenging for samplers to escape these trivial modes and explore the full Hilbert space.

To study the scaling of our proposed algorithm, we consider the cases $N\in\{30,40,50\}$ for which exact diagonalisation is out of reach for our computational resources. Since the Hamiltonian is real and we only study the time-independent ground state, we restrict ourselves to real-valued optimisation for our algorithm, although the structure carries over to complex cases \textit{mutatis mutandis}. 

\subsection{Details of Neural Network analysis}
 We implement our algorithm using \texttt{PyTorch (v2.7.0)}~\cite{10.5555/3454287.3455008} and \texttt{normflows}~\cite{Stimper2023} package. The NFs model is based on the Real Non Volume Preserving (RealNVP) flows~\cite{dinh2017density}.
This consists of affine coupling layers $\mathbf{z}_{i+1}=f_{i+1}(\mathbf{z}_i)$, where the $n$-dimensional vector $\mathbf{z}_i$ is divided into two vectors $\mathbf{z}_i=\mathbf{z}_i^{1:d}\oplus \mathbf{z}^{d+1:n}$, for some $d< n$. The two decomposed vectors undergo the operation
\begin{align}
\mathbf{z}_{i+1}^{1:d} &= \mathbf{z}_{i}^{1:d}\\
\mathbf{z}_{i+1}^{d+1:n} &= \mathbf{z}_{i}^{d+1:n} \odot \exp\big(s(\mathbf{z}_{i}^{1:d})\big) + t(\mathbf{z}_{i}^{1:d})\quad,
\end{align} where $\odot$ denotes the element-wise product, while $s(.)$ and $t(.)$ are learnable scaling and translation functions with $n-d$ dimensional outputs.
For an $N$-spin system, we set the posterior dimension to $n=N$ and apply a final $\tanh$ activation in our normalising flow. This confines \footnote{We bound the posterior's support for better training stability, even though the normally distributed continuous posterior has support on the entire space. Due to the small standard deviation of 0.1, this introduces an error of the order of $10^{-7}$ ($5\sigma$ from the mean 0.5), which is beyond the achievable precision with $N_{\mathrm{MC}}=25$.   } the posterior samples $\mathbf{y}$ to the hypercube $[-1,1]^N$. Each discrete spin configuration $\mathbf{x}$ is then obtained via the sign function defined in Eq.~\ref{eq:sign_function}. Consequently, each region $R_\mathbf{x}$ corresponds precisely to one orthant of the hypercube, facilitating straightforward computation of the integral volumes for Monte Carlo estimation as described in Eq.~\eqref{eq:mc_estimate}.

For all experiments, we use four coupling layers and set $d=\ceil{\frac{N}{2}}$. In each layer, a single dense neural network (DNN) with output dimensions $2(N-d)$ and two hidden layers of 512 nodes parametrises both the scaling and the translation functions.  This DNN has ReLU activation in the hidden layers and tanh output activation. The NQS model is also a DNN that takes in $n$-dimensional local quantum numbers of the basis and contains four hidden layers of 512 nodes and ReLU activation with a final tanh output activation. 

We construct the prior by independent sampling over each direction in $\mathbb{R}^n$ where each constituent sample consists of a mixture of two Gaussians located at $\pm 1$ with a standard deviation of $0.33$. For $N=10$, we sample subspaces of size $|S|=150$, while for $N\in\{15,20,...\}$, we consider $|S|=5000$. For the smoothing step, we sample $N_{\mathrm{MC}}=25$ samples within each region contained in $S$. The gradients are updated after $|B|=30$ such independent subspaces $S$ with the loss function $\mathcal{L}$ via the Adam optimiser with a learning rate of $10^{-4}$. 

\begin{figure}[ht!]
  \centering
  \includegraphics[scale=0.22]{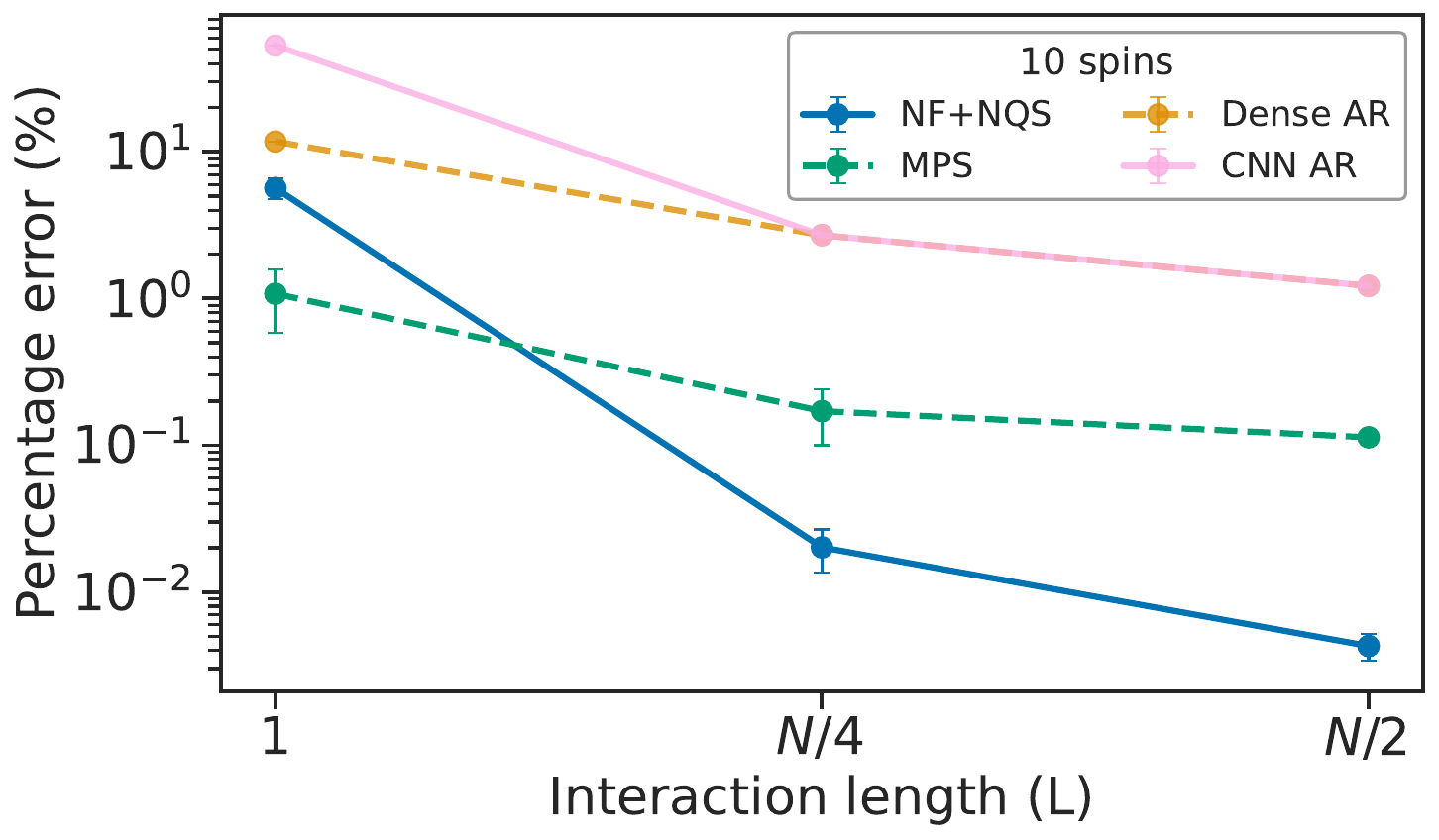}\\
\vspace{2pt}
  \includegraphics[scale=0.22]{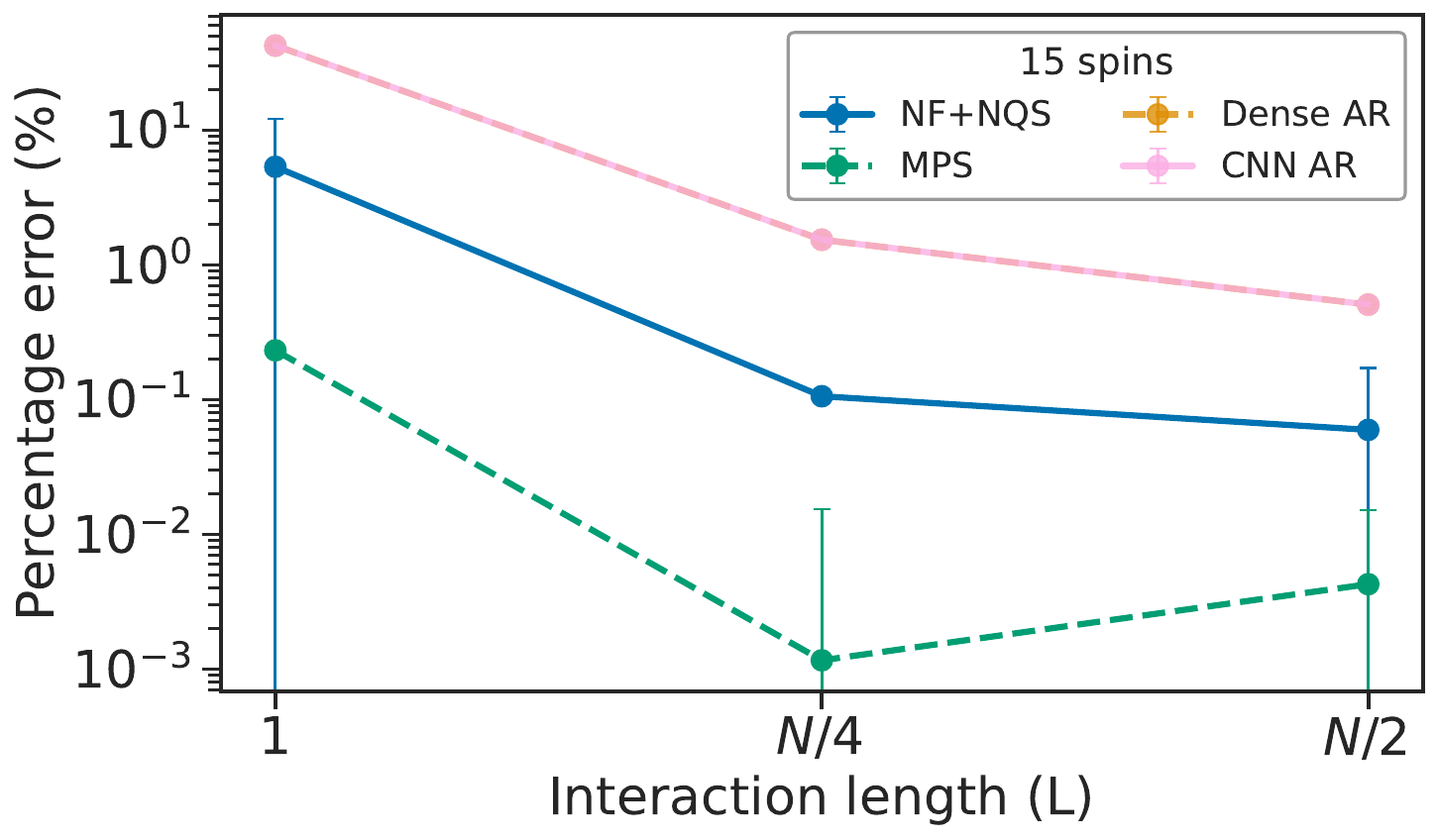}\\
\vspace{2pt}
   \includegraphics[scale=0.22]{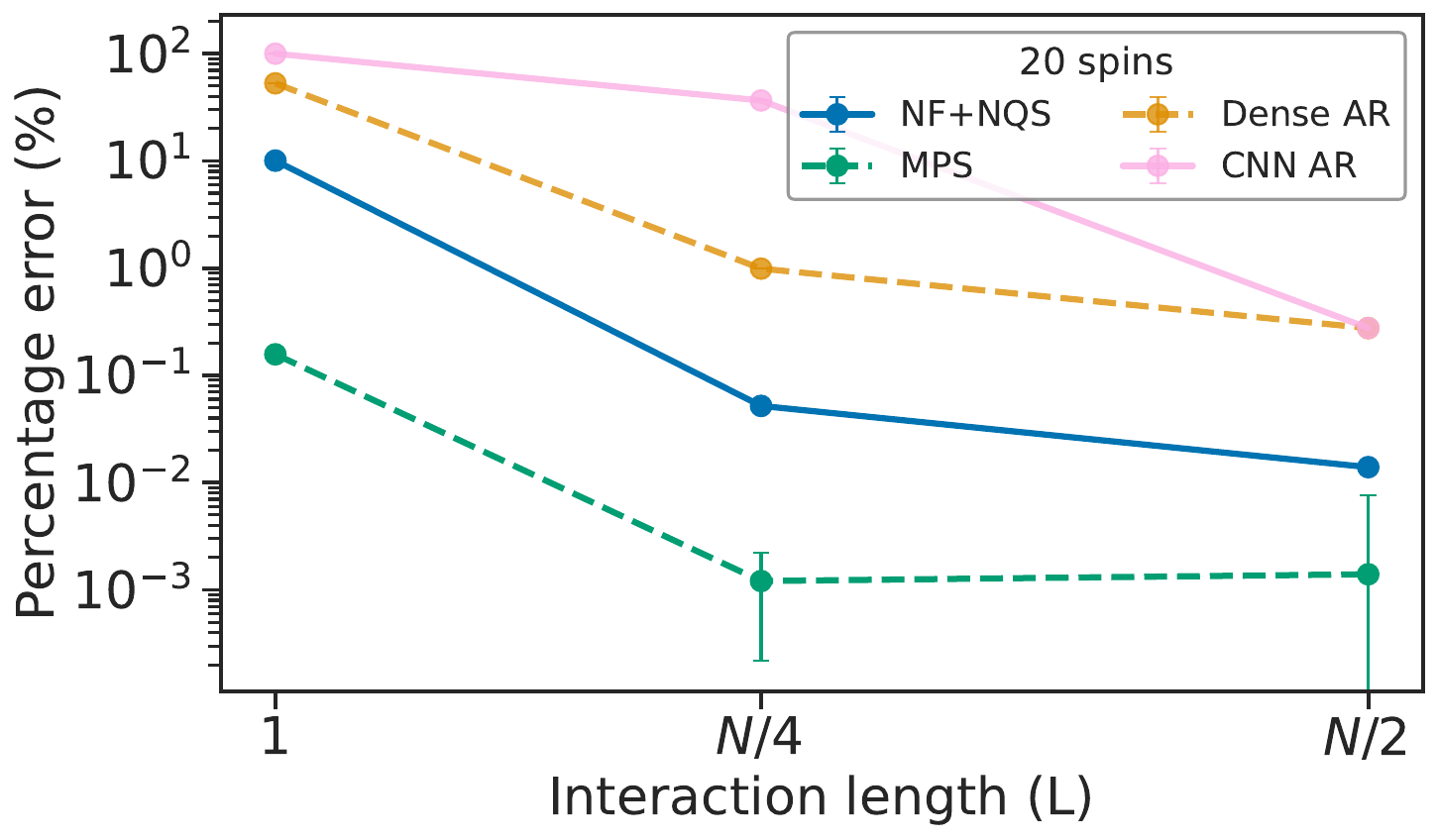}
  \caption{Percentage error for different interaction length ($L$) for $N\in\{10,15,20\}$ spin chains, with $V=1.0$.}
  \label{fig:spin_L_scan}
\end{figure}

To judge convergence, we let the training continue until the number of unique samples drop below $90 \%$ of $|S|$ in all sets $S$, after which we stop the training.  For our final inference, we sample $S$ samples from the converged NF model and train a randomly initialised NQS model with the same architecture as described above. 
It is trained for 2000 iterations, using the Adam optimiser with an initial learning rate of $10^{-3}$, which reduces by a factor of 0.5 if the energy expectation has not decreased for twenty steps.

We compare our method (NF+NQS) with Matrix Product State (MPS), Dense Autoregressive Model (Dense-AR) and one-dimensional Convolutional Autoregressive Model (CNN-AR) implemented in \texttt{NetKet} with just-in-time compilation based on \texttt{Numba}~\cite{lam2015numba} and  \texttt{Jax}~\cite{deepmind2020jax}. Note that our method utilises a strictly closed subspace in the energy expectation minimisation, while NetKet does a covariance minimisation over local energies. Thus, for $k$-dimensional basis set, we evaluate the energy expectation of the $k\times k$ Hamiltonian in the subspace while NetKet methods evaluates local energies of each basis in the set over the Hilbert space having an effective $k\times d^N$ matrix dimension. While our approach has less complexity, it utilises less information in every iteration.  

For MPS, we use the \texttt{MetropolisLocal} sampler, while for the autoregressive models, we use \texttt{ARDirectSampler}. We set the bond dimensions of the MPS model to be five for all experiments. The Dense-AR consists of six layers with each layer having 2N hidden nodes. The CNN-AR consists of four convolutional blocks, each with 512 kernels of size $\floor{\frac{N}{2}}$. To keep the number of samples approximately the same, we keep the total number of states in the  Monte Carlo state the same as those for our algorithm, i.e. 150 for N=10 and 5000 otherwise. All models are trained with the Adam optimiser, with an initial learning rate of 0.01 that decays by a factor of 0.5 every fifty iterations for a total of 400 iterations.

\begin{figure}[t]
  \centering
  \includegraphics[scale=0.22]{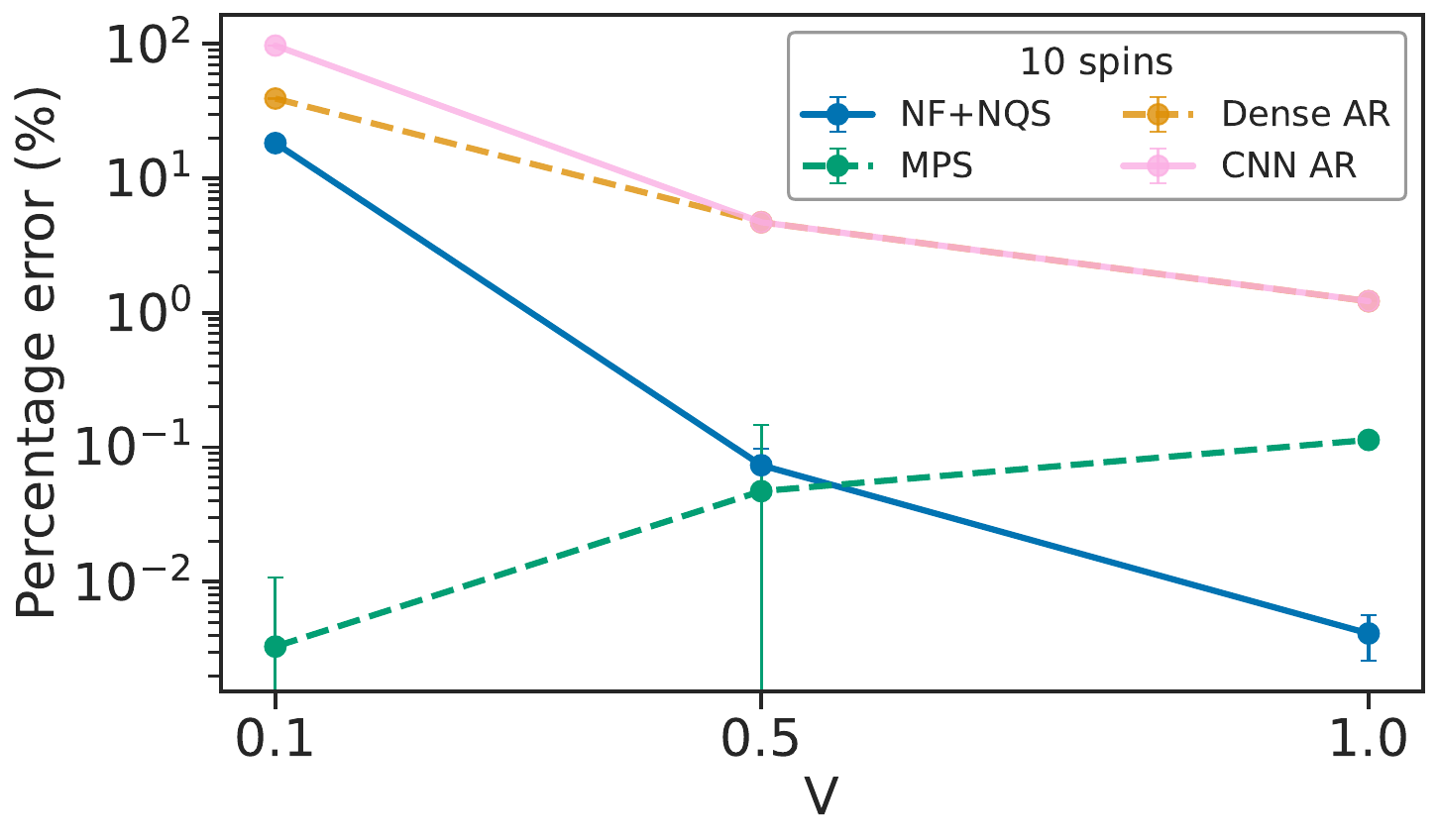}\\
\vspace{2pt}
  \includegraphics[scale=0.22]{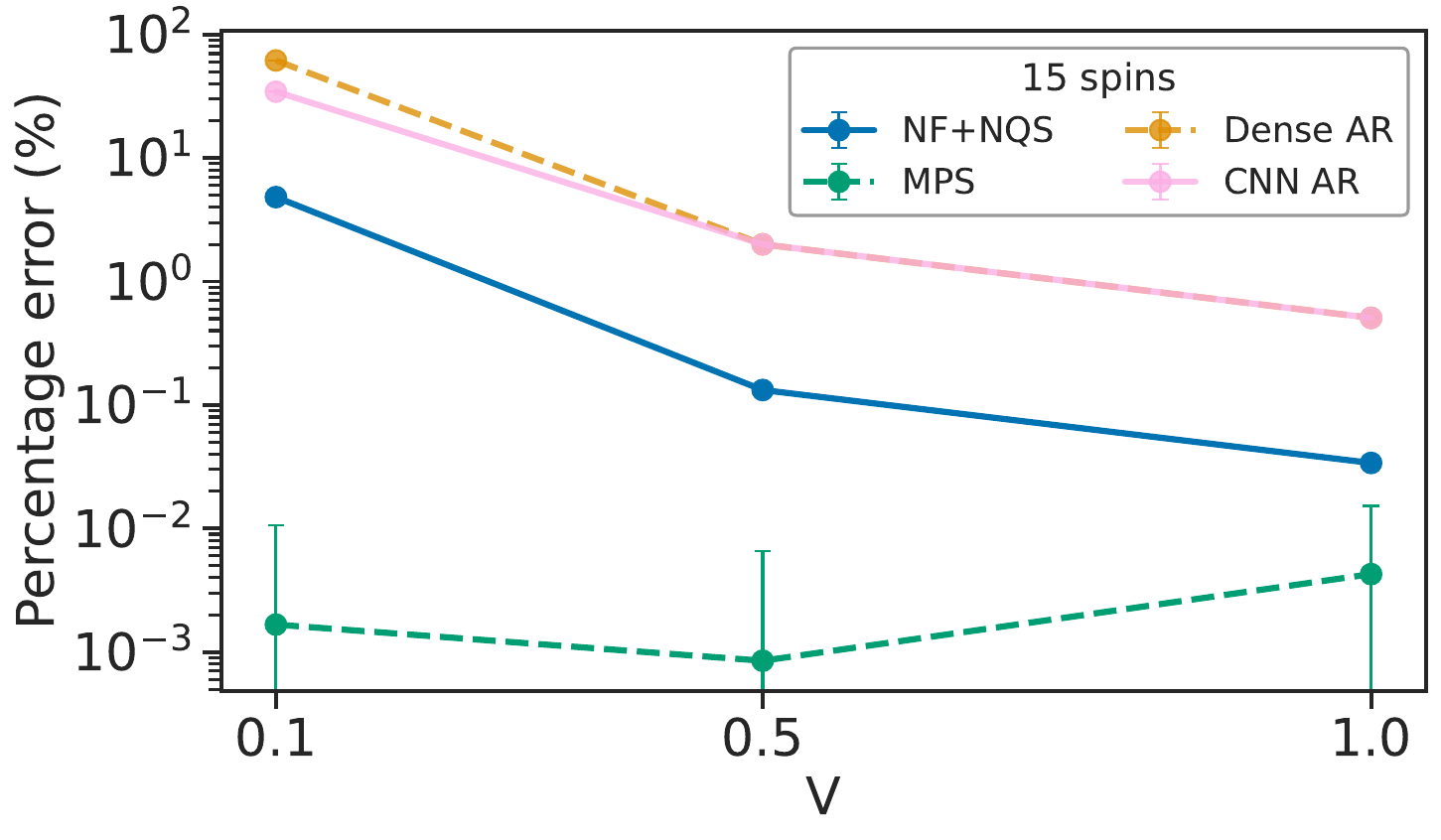}\\
\vspace{2pt}
  \includegraphics[scale=0.22]{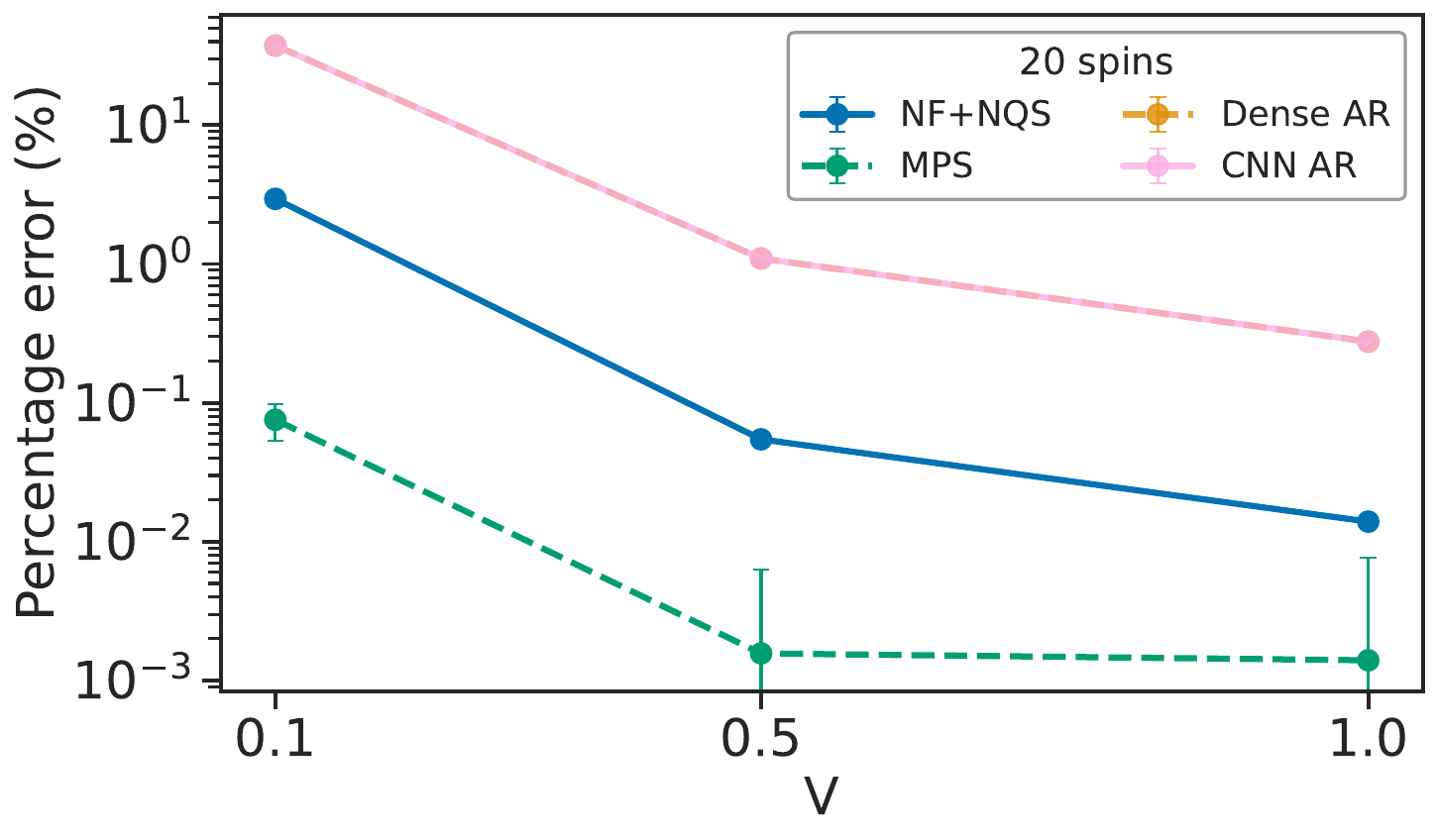}
  \caption{Percentage error for different interaction strengths ($V$) for $N\in\{10,15,20\}$ spin chains, with $L=\ceil{\frac{N}{2}}$.}
  \label{fig:spin_V_scan}
\end{figure}

\subsection{Results}

For systems with $N\in\{10,15,20\}$ spins, exact diagonalisation (ED) provided reference energies $E_{\rm true}$, which were used to construct the percentage error
\begin{equation}
  \mathrm{Error}\;[ \% ]
    \;=\;\frac{E_{\rm method}-E_{\rm true}}{|E_{\rm true}|}\times100\%.
\end{equation}
However, for larger system sizes ($N\in\{30,40,50\}$), ED became infeasible, and we report energies obtained directly from each method.

Error bars for NF+NQS results were estimated by sampling 20 independent subspaces using the NF model and minimising the energy expectation using a randomly initialised NQS. The mean energies are reported in the tables, with standard deviation representing uncertainty. CNN-AR and Dense-AR consistently became trapped predicting trivial configurations (all-up or all-down states), dominated by self-interactions within approximately 30 gradient updates, leading to zero variance in reported energies. This is the nature of the ferromagnetic phase, where the ground state has maximum probability to be in a spin-aligned state--the ground state displays modes in the all-up or all-down state. With increasing long-range interactions and or interaction strength, the relative importance of these two basis states increases. However, to find the quantum mechanical ground state, one needs to consider different basis states beyond these two, and we find our algorithm can learn to sample such states correctly.

Figure \ref{fig:spin_L_scan} presents percentage errors relative to ED ground state energies for chain sizes $N\in\{10,15,20\}$ as a function of interaction length ($L$). The NF+NQS method consistently demonstrated competitive accuracy across short ($L=1$), intermediate ($L=N/4$), and long ($L=N/2$) interaction regimes. Notably, accuracy generally improved with increasing interaction length, reflecting the enhanced capability of our NF-based model to capture non-local entanglement. For larger spin systems ($N\in\{30,40,50\}$), Table \ref{tab:spin_L_scan} illustrates that our NF+NQS method maintained robust performance in strongly entangled regimes (fully connected, $L=N/2$).


\begin{table}[t!]
  \centering
  \begin{tabular}{cc r@{\,±\,}l r@{\,±\,}l r@{\,±\,}l r@{\,±\,}l}
    \toprule
 $N$ &   $L$ & \multicolumn{2}{c}{NQS+NF} & \multicolumn{2}{c}{MPS} & \multicolumn{2}{c}{Dense AR} & \multicolumn{2}{c}{CNN AR} \\

\midrule
\multirow{ 3}{*}{30}  &  1 & -25.7 & 0.1 & -38.184 & 0.003 & -10 & 0 & -10 & 0 \\
  &  8 & -240.83 & 0.03 & -240.942 & 0.008 & -240 & 0 & -240 & 0 \\
  &  15 & -435.4 & 0.1 & -435.516 & 0.002 & -435 & 0 & -435 & 0 \\

\midrule
   \multirow{ 3}{*}{40} & 1 & -36.22 & 0.09 & -50.914 & 0.003 & -36 & 0 & -40 & 0 \\
   & 10 & -400.86 & 0.02 & -401.006 & 0.008 & -360 & 0 & -400 & 0 \\
   & 20 & -780.3 & 0.2 & -780.491 & 0.01 & -780 & 0 & -780 & 0 \\
    \midrule

 \multirow{ 3}{*}{50} &  1 & -38.89 & 0.08 & -63.643 & 0.004 & -18 & 0 & -26 & 0 \\
 &   13 & -650.3 & 0.2 & -650.956 & 0.004 & -650 & 0 & -650 & 0 \\
 &   25 & -1225.40 & 0.02 & -1225.24 & 0.07 & -1225 & 0 & -1225 & 0 \\
    \bottomrule
  \end{tabular}
  \caption{Ground state energies found via different methods for $N\in\{30,40,50\}$, and different interaction lengths $L$ keeping $V=1.0$. The error on the NQS+NF method is evaluated by minimising the energy expectation over 20 independent subspaces generated by the NF model with a randomly initialised NQS model.}
  \label{tab:spin_L_scan}
\end{table}


\begin{table}[h!]
\centering
\vspace{2pt} 
\begin{tabular}{c c r@{\,±\,}l r@{\,±\,}l r@{\,±\,}l r@{\,±\,}l}
\toprule
$N$ &$V$ & \multicolumn{2}{c}{NQS+NF} & \multicolumn{2}{c}{MPS} & \multicolumn{2}{c}{Dense AR} & \multicolumn{2}{c}{CNN AR} \\

\midrule
\multirow{3}{*}{30} &0.10 & -47.52 & 0.02 & -48.680 & 0.004 & -27.3 & 0 & -43.5 & 0 \\
&0.50 & -218.41 & 0.06 & -218.541 & 0.005 & -217.5 & 0 & -217.5 & 0 \\
&1.00 & -435.4 & 0.1 & -435.516 & 0.002 & -435 & 0 & -435 & 0 \\

\midrule
\multirow{3}{*}{40} &0.10 & -81.89 & 0.02 & -83.133 & 0.004 & -62.8 & 0 & -78 & 0 \\
&0.50 & -390.86 & 0.09 & -391.026 & 0.001 & -314 & 0 & -390 & 0 \\
&1.00 & -780.3 & 0.2 & -780.49 & 0.01 & -780 & 0 & -780 & 0 \\

\midrule
\multirow{3}{*}{50}& 0.10 & -126.15 & 0.04 & -127.606 & 0.002 & -85.7 & 0 & -122.5 & 0 \\
&0.50 & -613.0 & 0.3 & -613.50 & 0.02 & -428.5 & 0 & -612.5 & 0 \\
&1.00 & -1225.40 & 0.02 & -1225.24 & 0.07 & -1225 & 0 & -1225 & 0 \\
\bottomrule
\end{tabular}
\caption{Ground state energies found via different methods for $N\in\{30,40,50\}$, and different interaction strenths $V$, keeping $L=\ceil{\frac{N}{2}}$. The error on the NQS+NF method is evaluated by minimising the energy expectation over 20 independent subspaces generated by the NF model with a randomly initialised NQS model.}
\label{tab:spin_V_scan}
\end{table}

Similarly, Figure \ref{fig:spin_V_scan} show percentage error across different interaction strengths ($V\in\{0.1,0.5,1.0\}$) for spin chain sizes $N\in\{10,15,20\}$. The NQS+NF method demonstrated strong reliability across varying interaction strengths, excelling particularly at strong coupling ($V=1.0$), which is a challenging regime due to significant entanglement. Table \ref{tab:spin_V_scan} summarises energies obtained at varying interaction strengths for larger systems ($N=30,40,50$), further highlighting the method's consistent performance, particularly at higher interaction strengths.

\section{Summary and Conclusions} 
\label{sec:conclusion}

In this work, we proposed a hybrid variational framework that augments Neural Quantum States with a Normalising Flow-based sampler to improve ground state estimation in quantum many-body systems. By decoupling the sampling process from the variational ansatz, our method overcomes fundamental limitations of MCMC and autoregressive techniques, particularly in the presence of volume-law entanglement and long-range correlations. The key idea lies in using continuous normalising flows to sample efficiently from a discretised effective subspace of the Hilbert space, enabling expressive and non-local exploration of quantum configurations that are otherwise inaccessible to sequential or local samplers.

We validated our approach on the transverse-field Ising model with varying interaction ranges and strengths. Across system sizes $N=10$ to $N=50$, our NF+NQS framework consistently achieved lower or comparable ground state energy errors relative to benchmark methods such as matrix product states and autoregressive networks. For smaller systems where exact diagonalisation was possible, we observed percentage energy errors below $1\%$ even in highly entangled regimes. For larger systems ($N=30$–$50$), our method maintained competitive accuracy while autoregressive models collapsed into trivial configurations. Notably, we demonstrated the robustness of our method across both weakly and strongly coupled regimes and for fully connected interaction graphs.

These results suggest the promise of flow-assisted sampling as a scalable and accurate tool for quantum simulation. In future work, this framework can be extended to complex-valued wavefunctions and fermionic systems, or combined with gauge-invariant architectures for lattice gauge theories. Additionally, the decoupled sampling strategy opens the door to using more expressive deep generative models beyond normalising flows, including diffusion models or transformer-based samplers. Overall, our work contributes a flexible and powerful tool for quantum state learning, with potential improvements in a wide range of quantum field theory applications.
	
\bibliographystyle{apsrev4-1}
\bibliography{ref.bib}

\clearpage

\end{document}